\pdfoutput=1
\documentclass[letterpaper,10pt,conference]{ieeeconf}
\IEEEoverridecommandlockouts
\usepackage[T1]{fontenc}
\usepackage{amsmath,amssymb}
\allowdisplaybreaks
\usepackage{bm}
\usepackage{cite}
\usepackage{graphicx}
\graphicspath{{./}{ECC/}}
\usepackage{booktabs}
\usepackage{placeins}
\usepackage{stfloats}
\usepackage{algorithm}
\usepackage{algpseudocode}

\makeatletter\g@addto@macro\fs@ruled{\def\@fs@cfont{\rmfamily}}\makeatother
\usepackage[hidelinks]{hyperref}
\usepackage{xcolor}

\newtheorem{assumption}{Assumption}
\newtheorem{lemma}{Lemma}
\newtheorem{theorem}{Theorem}

\newcommand{\col}{\operatorname{col}}
\newcommand{\norm}[1]{\left\lVert#1\right\rVert}
\newcommand{\qedsq}{\hfill$\square$\par}
\newcommand{\proofhead}{\noindent\hspace{2em}{\itshape Proof: }}

\title{ Information-Region MPC:\\ Non-Iterative Distributed Control of Urban Traffic Signals}
\author{Qiongdan Hu$^{1}$, Menglin Yang$^{1}$, Meng Wang$^{1}$, and Chaopeng Tan$^{1}$
\thanks{This work was funded by the Deutsche Forschungsgemeinschaft (DFG, German Research Foundation) -- TRR 408/1 2025 -- Project number 531327426.}
\thanks{$^{1}$The authors are with the Chair of Traffic Process Automation, Technische Universit\"at Dresden, 01069 Dresden, Germany.\protect\\ {\tt\scriptsize qiongdan.hu@tu-dresden.de, menglin.yang@tu-dresden.de, meng.wang@tu-dresden.de, chaopeng.tan@tu-dresden.de}}%
\thanks{The code is available at\protect\\ {\scriptsize\protect\url{https://github.com/Tarard/Information-Region-MPC}}.}%
}

\newcommand{\dispgap}{3.5pt plus 1pt minus 1pt}
\AtBeginDocument{\setlength{\abovedisplayskip}{\dispgap}\setlength{\belowdisplayskip}{\dispgap}\setlength{\abovedisplayshortskip}{\dispgap}\setlength{\belowdisplayshortskip}{\dispgap}}
\begin{document}
\maketitle
\thispagestyle{empty}
\pagestyle{empty}

\begin{abstract}
Centralized model predictive control (MPC) exploits network-wide traffic information to obtain the optimal solution of the network-level control problem, but its computational burden grows rapidly with network size. Iterative distributed MPC reduces the size of local optimization problems, but requires repeated information exchange and coordination within each control update. To address these limitations, this paper proposes information-region MPC (IR-MPC), which approximates the first control input of centralized MPC using limited real-time data without iterative coordination. Based on historical traffic data, we quantify how perturbations at other intersections affect the centralized first control input of each intersection and use these couplings to construct a compact information region for every intersection. At each control update, each intersection then solves a local IR-MPC problem once using current data only from its information region. SUMO simulations on a real 28-intersection network show that IR-MPC with one processor per intersection computes its control inputs 9 times faster than centralized MPC and 15 times faster than a representative iterative decomposition-based distributed MPC method. IR-MPC reproduces 85\% of the centralized control inputs exactly at the implemented one-second resolution, keeps 99.3\% within 2~s, and increases mean delay by only 0.03\%.
\end{abstract}

\section{Introduction}
Model predictive control (MPC) is an established method for coordinating the signals of an urban network~\cite{aboudolas2010,lin2012}. Centralized MPC uses network-wide traffic information to optimize the green times of all intersections jointly and thus obtains the optimal solution of the network-level MPC problem. However, this joint optimization becomes increasingly computationally demanding as the network grows. Distributed and hierarchical MPC reduce this burden in two different ways~\cite{scattolini2009}. In iterative distributed MPC, each intersection solves a local problem, and neighboring controllers exchange predicted flows or prices until their solutions agree, for example through dual decomposition~\cite{giselsson2013}, the alternating direction method of multipliers (ADMM)~\cite{pham2023}, or overlapping Schwarz decomposition (OSD)~\cite{shin2020schwarz}. Each iteration of exchange takes time, and many iterations can be necessary within one signal update before the local solutions come close to the centralized one. Hierarchical MPC avoids these iterations: an upper layer optimizes the flows between regions with an aggregated model such as the macroscopic fundamental diagram, and the local controllers follow its targets~\cite{ramezani2015}. Because the upper layer is coarser than the local models, the result no longer approximates centralized MPC on the detailed model. Max-pressure control is also non-iterative and decentralized, but each intersection uses only the information of its neighborhood~\cite{mercader2020}, although strong couplings can also occur between non-adjacent intersections.

\begin{figure}[!t]
\centering
\includegraphics[width=\columnwidth]{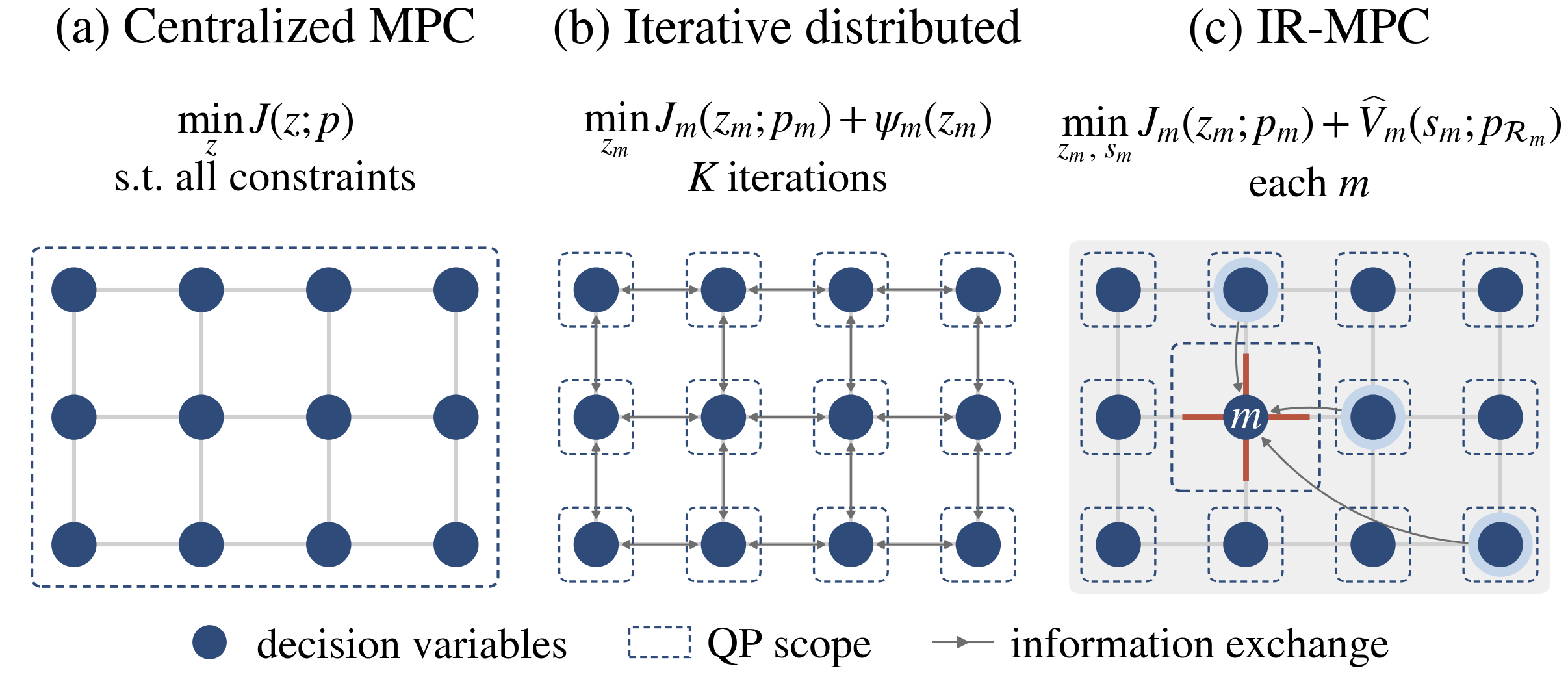}
\caption{QP scope and information exchange in centralized MPC, iterative distributed MPC with coordination term $\psi_m$ and $K$ iterations, and IR-MPC.}
\label{fig:paradigms}
\end{figure}

To bridge this gap, we propose information-region MPC (IR-MPC), a non-iterative distributed control framework that approximates the first control input of centralized MPC. For each intersection, the influence of the remaining network is represented by a boundary value function defined over the traffic flows exchanged between the local intersection and the rest of the network. Historical centralized MPC solutions are used offline to approximate it and to quantify the sensitivity of the centralized control input at each intersection to traffic-state variations at other intersections. These sensitivities characterize the coupling between intersections and are used to construct an information region for each local controller. During operation, each intersection acquires current traffic data only from its information region, updates this approximation accordingly, and solves one local quadratic program (QP) to determine its next control input, without iterative coordination among controllers. We bound the deviation of the local first control input from that of centralized MPC by the error of this approximation plus the couplings to the intersections outside the information region, which provides a basis for selecting the information region for a prescribed control-approximation tolerance. Figure~\ref{fig:paradigms} compares the optimization scope and information exchange of centralized MPC, iterative distributed MPC, and IR-MPC.

The contributions of this paper are twofold. First, we introduce an information-region paradigm for distributed MPC, in which each intersection approximates the first control input of centralized MPC by solving a single local QP using current traffic data only from its information region, without iterative coordination among local controllers. Second, we develop a sensitivity-based method to construct the information regions by quantifying how traffic-state variations at other intersections affect the centralized first control input. We further derive an error bound that relates the omitted inter-intersection couplings to the resulting control-input deviation, providing a principled basis for selecting the information region, and show that strong couplings do not necessarily follow network hop distance. Simulations of a real 28-intersection network examine whether the couplings yield compact information regions and compare the control-input accuracy, closed-loop traffic performance, and computation time of IR-MPC with those of centralized MPC and OSD. Synthetic grids with up to 500 intersections test how the computation and information requirements scale with network size.

\section{Problem Statement}\label{sec:problem}

This section defines the centralized MPC benchmark and the local approximation problem solved by IR-MPC. We first formulate the network-level MPC as a convex parametric QP, and then expose its intersection-wise structure to define the information-region controller. Section~\ref{sec:irmpc} constructs the approximation required by the local controller.

\subsection{Centralized MPC}

We first establish the network-level control problem whose first control input IR-MPC seeks to reproduce. At each control update, centralized MPC uses network-wide traffic measurements to predict the network evolution over a finite horizon and jointly optimize future signal timings.

Let $\mathcal V$ and $\mathcal Z$ denote the sets of signalized intersections and links, respectively, and let $\mathcal Z_{\mathrm{in}}\subseteq\mathcal Z$ be the set of network-entry links. Time is discrete with step length $T$, and $t$ denotes the current control update. We use the store-and-forward model~\cite{aboudolas2009,aboudolas2010} with the state $\bm x(t)=\col(\bm n(t),\bm w(t))\in\mathbb R^{n_x}$, the flow vector $\bm f(t)=\col(\bm f^{\mathrm{out}}(t),\bm f^{\mathrm{in}}(t))\in\mathbb R^{n_x}$, and the disturbance $\bm d(t)\in\mathbb R^{|\mathcal Z_{\mathrm{in}}|}$, where $n_x=|\mathcal Z|+|\mathcal Z_{\mathrm{in}}|$. Here $\bm n$ and $\bm f^{\mathrm{out}}$ are the link-accumulation and link-outflow vectors, $\bm w$ and $\bm f^{\mathrm{in}}$ are the entry-queue and admitted-entry-flow vectors, and $\bm d$ is the entry-arrival vector. The control input $\bm u\in\mathbb R^{n_u}$ stacks $\bm u_{m,\nu}\in\mathbb R^{n_{\phi,m}}$, the green times of the $n_{\phi,m}$ phases of intersection $m$ in its $\nu$th future cycle, for $\nu=1,\ldots,N_m$ and $m\in\mathcal V$, where $N_m$ is the number of cycles of $m$ that start within the horizon. The control input limits the outflows through~\eqref{eq:service}. The prediction horizon contains $N$ steps. The data vector $\bm p$ contains the measured state $\bm x(t)$, the estimated arrival rates $\hat{\bm d}(t)$, and the green times of the most recently started cycles.

At update time $t$, centralized MPC solves
\begin{subequations}\label{eq:ocp}
\begin{align}
\min_{\rlap{$\scriptstyle\bm u,\bm f,\bm x$}}\quad&J_x+J_g+J_R\label{eq:ocp-obj}\\
&J_x=\tfrac12\textstyle\sum_{k=1}^{N-1}\|\bm x(t{+}k)\|_{\bm W_x}^2+\tfrac12\|\bm x(t{+}N)\|_{\bm P}^2,\label{eq:cost-state}\\
&J_g=\tfrac{\mu_g}{2}\textstyle\sum_{m\in\mathcal V}\sum_{\nu=1}^{N_m}\bigl\|\bm u_{m,\nu}-\bm u_{m,\nu-1}\bigr\|_2^2,\label{eq:cost-green}\\
&J_R=\tfrac{\mu_R}{2}\textstyle\sum_{m\in\mathcal V}\sum_{\nu=1}^{N_m}\|\bm r_{m,\nu}\|_{\bm W_{R,m,\nu}}^2,\label{eq:cost-delay}\\
\text{s.t.}\quad&\bm x(t{+}k{+}1)=\bm x(t{+}k)+\bm B\bm f(t{+}k)+\bm B_d\bm d(t{+}k),\nonumber\\
&\bm d(t{+}k)=\hat{\bm d}(t),\quad\bm f^{\mathrm{in}}(t{+}k)\ge\bm 0,\label{eq:saf}\\
&\bm 0\le\bm f^{\mathrm{out}}(t{+}k)\le\bm S\bm\Phi\bigl[\bm R_k\bm u+\bm q_k(\bm p)\bigr],\label{eq:service}\\
&\bm 1^\top\bm u_{m,\nu}=C_m-L_m,\quad\underline{\bm u}_m\le\bm u_{m,\nu}\le\overline{\bm u}_m,\label{eq:green}\\
&\bm 0\le\bm n(t{+}k)\le\overline{\bm n},\quad\bm w(t{+}k)\ge\bm 0,\label{eq:storage}\\
&\bm x(t)\text{ is measured}.\label{eq:init}
\end{align}
\end{subequations}
The dynamics and service constraints apply for $k=0,\ldots,N-1$, the storage constraints for $k=1,\ldots,N$, and the green-time constraints for $\nu=1,\ldots,N_m$ and $m\in\mathcal V$.

The cost function has three components. $J_x$ penalizes link accumulations and entry queues with the stage weight $\bm W_x=\operatorname{diag}(\bm W_n,\bm W_w)$ and the terminal weight $\bm P=\bm W_x$, where the diagonal matrices $\bm W_n$ and $\bm W_w$ normalize their scales and $\|\bm x\|_{\bm W}^2=\bm x^\top\bm W\bm x$. $J_g$ regularizes cycle-to-cycle green-time changes, where $\bm u_{m,0}$, a part of $\bm p$, contains the green times of the most recently started cycle of $m$. $J_R$ is a movement-level delay term based on Webster's uniform-delay expression~\cite{webster1958}. In $J_R$, $\bm r_{m,\nu}=C_m\bm 1-\bm A_{R,m}\bm u_{m,\nu}$ contains the effective red times of the movements of $m$, where $\bm A_{R,m}$ maps phase green times to movement green times, and the diagonal matrix $\bm W_{R,m,\nu}$ contains the movement weights and the share of cycle $\nu$ inside the horizon. $J_g$ and $J_R$ sum over the signal cycles in the horizon, because one cycle can cover more than one step. The positive coefficients $\mu_g$ and $\mu_R$ control the latter two terms.

Constraints~\eqref{eq:saf}--\eqref{eq:init} describe network conservation and entry queues, signal service, cycle-time limits, storage up to the capacities $\overline{\bm n}$, and the measured initial state. In~\eqref{eq:saf}, $\bm B=T\left[\begin{smallmatrix}\bm\Gamma^\top-\bm I&\bm B_{\mathrm{in}}\\\bm 0&-\bm I\end{smallmatrix}\right]$ and $\bm B_d=T\left[\begin{smallmatrix}\bm 0\\\bm I\end{smallmatrix}\right]$, where $\bm\Gamma\in\mathbb R^{|\mathcal Z|\times|\mathcal Z|}$ is the turning-ratio matrix and $\bm B_{\mathrm{in}}\in\mathbb R^{|\mathcal Z|\times|\mathcal Z_{\mathrm{in}}|}$ maps admitted entry flows to links. The vector $\hat{\bm d}(t)$ contains the estimated entry-arrival rates, held over the horizon. In~\eqref{eq:service}, $\bm S\in\mathbb R^{|\mathcal Z|\times|\mathcal Z|}$ is the diagonal matrix of saturation flows, $\bm\Phi\in\mathbb R^{|\mathcal Z|\times n_\phi}$ is the phase--link service matrix, and $\bm R_k\bm u+\bm q_k(\bm p)\in\mathbb R^{n_\phi}$ gives the green splits of all $n_\phi$ phases at step $k$. In~\eqref{eq:green}, $C_m$ and $L_m$ are the cycle length and lost time of $m$, and $\underline{\bm u}_m$ and $\overline{\bm u}_m$ contain the minimum and maximum green times.

Following the simultaneous solution concept for optimal control problems, we define the decision vector $\bm z$ that stacks $\bm x(t)$, $\bm f(t+k)$, and $\bm x(t+k+1)$ for $k=0,\ldots,N-1$ together with $\bm u$. Because~\eqref{eq:ocp} has a quadratic cost function and affine constraints, it can then be recast into the convex parametric QP
\begin{equation}\label{eq:full}
\begin{aligned}
J^\star(\bm p)
=\min_{\bm z}\quad&
J(\bm z;\bm p)
:=
\tfrac12\bm z^\top\bm H\bm z+
\bm h(\bm p)^\top\bm z\\
\mathrm{s.t.}\quad&
\bm E\bm z=\bm e(\bm p),\\
&
\bm G\bm z\le\bm g(\bm p),
\end{aligned}
\end{equation}
where $\bm z\in\mathbb R^{n_z}$, $\bm H=\bm H^\top\in\mathbb R^{n_z\times n_z}$ is positive semidefinite, and $\bm E\in\mathbb R^{n_e\times n_z}$ and $\bm G\in\mathbb R^{n_i\times n_z}$ are independent of $\bm p$. The maps $\bm h$, $\bm e$, and $\bm g$ are affine in $\bm p$; for example, $J_g$ gives the term $-\mu_g\bm u_{m,0}^\top\bm u_{m,1}$, which is linear in $\bm z$ with a coefficient affine in $\bm p$. The terms $\tfrac12\bm x(t)^\top\bm W_x\bm x(t)$ and $\tfrac{\mu_g}{2}\|\bm u_{m,0}\|_2^2$ depend only on $\bm p$; they are omitted because they do not change the optimizer. Hence any optimizer $\bm z^\star(\bm p)$ is globally optimal for the formulated centralized MPC problem. For intersection $m$, let $\mathcal I_m$ index $\bm u_{m,1}$ in $\bm z$ and in $\bm z_m$. Its centralized first control input is $\bm a_m^\star(\bm p):=[\bm z^\star(\bm p)]_{\mathcal I_m}$, the green times of its next cycle.

\subsection{IR-MPC formulation}

Building on~\eqref{eq:full}, we now define the local problem solved by each intersection. Each link is assigned to the intersection controlling its outflow, so the variables, online data, and objective can be partitioned as
\[
\begin{aligned}
\bm z&=\col_{m\in\mathcal V}\bm z_m,
\qquad
\bm p=\col_{m\in\mathcal V}\bm p_m,\\
J(\bm z;\bm p)
&=\sum_{m\in\mathcal V}J_m(\bm z_m;\bm p_m).
\end{aligned}
\]
Under this partition, local signal and bound constraints involve one intersection, while inter-intersection coupling enters through the flow-conservation equations~\eqref{eq:saf}.

For intersection $m$, let $\mathcal R_m\subseteq\mathcal V$ be its information region, with $m\in\mathcal R_m$, and let $\mathcal O_m:=\mathcal V\setminus\{m\}$ denote the remaining intersections. The vector $\bm s_m$ stacks the inter-intersection flows coupling $m$ to $\mathcal O_m$ over the prediction horizon. IR-MPC represents the optimal response of $\mathcal O_m$ through an approximate boundary value function $\widehat V_m$ and solves
\begin{equation}\label{eq:irmpc}
\begin{aligned}
\min_{\bm z_m,\bm s_m}\quad&
J_m(\bm z_m;\bm p_m(t))
+
\widehat V_m
\bigl(\bm s_m;\bm p_{\mathcal R_m}(t)\bigr)\\
\mathrm{s.t.}\quad&
(\bm z_m,\bm s_m)
\in\mathcal X_m(\bm p_m(t)),\\
&
\bm s_m\in\mathcal S_m(t),
\end{aligned}
\end{equation}
where $\bm p_{\mathcal R_m}(t)$ stacks the current data in $\mathcal R_m$, $\mathcal X_m$ contains the local constraints inherited from~\eqref{eq:full}, and $\mathcal S_m(t)$ is the admissible set of boundary flows. Current data outside $\mathcal R_m$ are replaced by historical references in $\widehat V_m$.

Let $\bm z_m^{\mathcal R_m}(t)$ be the $\bm z_m$-part of an optimizer of~\eqref{eq:irmpc}. The implemented control is $\bm a_m^{\mathcal R_m}(t):=[\bm z_m^{\mathcal R_m}(t)]_{\mathcal I_m}$. Section~\ref{sec:irmpc} constructs $\widehat V_m$, and Section~\ref{sec:bounds} shows how the information excluded from $\mathcal R_m$ affects the deviation from $\bm a_m^\star(\bm p(t))$.

\section{Boundary Value Function Approximation}\label{sec:irmpc}

This section constructs $\widehat V_m$ in~\eqref{eq:irmpc}. The idea is to separate the centralized QP exactly at intersection $m$, represent the remaining network by its optimal value as a function of the exchanged flows, and then obtain a quadratic local representation of this boundary value function from historical centralized solutions. During online operation, only the linear term of this representation is updated using current data from $\mathcal R_m$.

\subsection{Primal decomposition}

We first make explicit the quantity that $\widehat V_m$ must represent. For intersection $m$, let
$\bm z_{\mathcal O_m}=\col_{j\in\mathcal O_m}\bm z_j$ and
$\bm p_{\mathcal O_m}=\col_{j\in\mathcal O_m}\bm p_j$.
Let $\bm s_m=\bm E_{m,\mathcal O_m}\bm z_{\mathcal O_m}$ collect the inflows that $m$ receives from its neighbors and the outflows that it sends to them in every prediction step, where $\bm E_{m,\mathcal O_m}$ has full row rank and the same row space as the $\bm z_{\mathcal O_m}$-block of the continuity equations~\eqref{eq:saf} that contain $\bm z_m$. With this $\bm s_m$, standard primal decomposition~\cite{conejo2006,boyd2015decomposition} separates the constraints of~\eqref{eq:full} into two sets. The set $\mathcal X_m(\bm p_m)$ contains the pairs $(\bm z_m,\bm s_m)$ that satisfy the constraints involving $\bm z_m$, in which $\bm s_m$ gives the exchanged flows in the continuity equations. The set $\mathcal F_m(\bm s_m;\bm p_{\mathcal O_m})$ contains the responses $\bm z_{\mathcal O_m}$ of the remaining network that satisfy the other constraints, $\bm E_{\mathcal O_m}\bm z_{\mathcal O_m}+\bm F_{\mathcal O_m}\bm s_m=\bm e_{\mathcal O_m}(\bm p_{\mathcal O_m})$ and $\bm G_{\mathcal O_m}\bm z_{\mathcal O_m}\le\bm g_{\mathcal O_m}(\bm p_{\mathcal O_m})$, where these matrices and vectors are blocks of~\eqref{eq:full} and $\bm F_{\mathcal O_m}\bm s_m$ equates the flows of the remaining network to $\bm s_m$. Thus $\bm z$ is feasible for~\eqref{eq:full} if and only if $(\bm z_m,\bm s_m)\in\mathcal X_m(\bm p_m)$ and $\bm z_{\mathcal O_m}\in\mathcal F_m(\bm s_m;\bm p_{\mathcal O_m})$ for some $\bm s_m$.

Because the objective is separable by intersection, the optimal cost of the remaining network is the boundary value function
\begin{equation}\label{eq:value}
\begin{aligned}
V_m(\bm s_m;\bm p_{\mathcal O_m})
=
\inf_{\bm z_{\mathcal O_m}}\quad&
J_{\mathcal O_m}
(\bm z_{\mathcal O_m};\bm p_{\mathcal O_m})\\
\mathrm{s.t.}\quad&
\bm z_{\mathcal O_m}\in
\mathcal F_m(\bm s_m;\bm p_{\mathcal O_m}),
\end{aligned}
\end{equation}
where
$J_{\mathcal O_m}
=\tfrac12\bm z_{\mathcal O_m}^{\top}\bm H_{\mathcal O_m}\bm z_{\mathcal O_m}
+\bm h_{\mathcal O_m}(\bm p_{\mathcal O_m})^\top\bm z_{\mathcal O_m}$ uses the blocks $\bm H_{\mathcal O_m}$ and $\bm h_{\mathcal O_m}$ of $\bm H$ and $\bm h$,
and $V_m=+\infty$ when the feasible set is empty. Hence~\eqref{eq:full} is equivalent to
\begin{equation}\label{eq:reduction}
\begin{aligned}
J^\star(\bm p)
=
\min_{\bm z_m,\bm s_m}\quad&
J_m(\bm z_m;\bm p_m)
+
V_m(\bm s_m;\bm p_{\mathcal O_m})\\
\mathrm{s.t.}\quad&
(\bm z_m,\bm s_m)\in\mathcal X_m(\bm p_m)
\end{aligned}
\end{equation}
in the variables $\bm z_m$ and $\bm s_m$, so replacing the optimization over $\mathcal O_m$ by $V_m$ is exact and the approximation in IR-MPC enters only through the approximation of $V_m$.

\subsection{KKT analysis at historical references}

We next construct that representation around a historical operating point. For update time $t$, let
$\bar{\bm p}(t)=\col_{m\in\mathcal V}\bar{\bm p}_m(t)$,
where $\bar{\bm p}_m(t)$ is the historical mean of $\bm p_m(t)$.
Let
$\Omega=\prod_{m\in\mathcal V}\Omega_m$
be a convex polytope of the observed data containing $\bar{\bm p}(t)$, and assume that the feasible set of~\eqref{eq:full} is nonempty and compact for every $\bm p\in\Omega$.
Solving~\eqref{eq:full} at $\bar{\bm p}(t)$ gives an optimizer
$\bm z^\star(\bar{\bm p}(t))$ and the associated reference exchange
$\bm s_m^0(t)$. The admissible set $\mathcal S_m(t)$ is defined by componentwise lower and upper bounds and contains this reference.

\begin{assumption}\label{as:coverage}
For each $\bm p\in\Omega$, at least one optimizer $\bm z^\star$ of~\eqref{eq:full} has its boundary flow $\bm s_m^\star$ in $\mathcal S_m(t)$.
\end{assumption}

Consider~\eqref{eq:value} at
$(\bm s_m^0(t),\bar{\bm p}_{\mathcal O_m}(t))$.
Let $\mathcal A_m^0$ index a selected subset of active inequalities, and let
$\bm\pi_m$ and $\bm\lambda_m$ be the multipliers of the equality and selected active inequality constraints, respectively. The corresponding KKT conditions are
\begin{equation}\label{eq:kkt}
\bm K_m(t)
\begin{bmatrix}
\bm z_{\mathcal O_m}\\
\bm\pi_m\\
\bm\lambda_m
\end{bmatrix}
=
\begin{bmatrix}
-\bm h_{\mathcal O_m}(\bm p_{\mathcal O_m})\\
\bm e_{\mathcal O_m}(\bm p_{\mathcal O_m})
-\bm F_{\mathcal O_m}\bm s_m^0(t)\\
\bm g_{\mathcal O_m,\mathcal A_m^0}
(\bm p_{\mathcal O_m})
\end{bmatrix},
\end{equation}
with
\begin{equation}\label{eq:kktmatrix}
\bm K_m(t)=
\begin{bmatrix}
\bm H_{\mathcal O_m}
&
\bm E_{\mathcal O_m}^{\top}
&
\bm G_{\mathcal O_m,\mathcal A_m^0}^{\top}\\
\bm E_{\mathcal O_m}&\bm 0&\bm 0\\
\bm G_{\mathcal O_m,\mathcal A_m^0}&\bm 0&\bm 0
\end{bmatrix}.
\end{equation}

\begin{lemma}\label{lem:nominal}
If $\bm E_{\mathcal O_m}$ has full row rank, then at
$(\bm s_m^0(t),\bar{\bm p}_{\mathcal O_m}(t))$
there exist an optimizer of~\eqref{eq:value} and a subset
$\mathcal A_m^0$ of its active inequalities with nonnegative KKT multipliers for which $\bm K_m(t)$ in~\eqref{eq:kktmatrix} is nonsingular.
\end{lemma}

\proofhead
Take an extreme point of the compact optimal set, and let $\mathcal D$ be the null space of $\bm E_{\mathcal O_m}$ and the inequality rows active at this point. $\bm H_{\mathcal O_m}$ is positive definite on $\mathcal D$, since a nonzero direction in $\mathcal D$ with zero curvature would, by KKT stationarity, generate an optimal segment and contradict extremality. Farkas' lemma and conic Carath\'eodory's theorem on $\ker\bm E_{\mathcal O_m}$~\cite[Props.~2.3.1, 1.2.1]{bertsekas2009} give linearly independent active rows with nonnegative multipliers. Adding active rows with zero multipliers until these rows and $\bm E_{\mathcal O_m}$ span all active rows gives a full-row-rank matrix with null space $\mathcal D$, so $\bm K_m(t)$ is nonsingular~\cite[Lemma~16.1]{nocedal2006}.\qedsq

Fix the active set from Lemma~\ref{lem:nominal}, and let
$\bm z_{\mathcal O_m}^0(\bm p_{\mathcal O_m})$,
$\bm\pi_m^0(\bm p_{\mathcal O_m})$, and
$\bm\lambda_m^0(\bm p_{\mathcal O_m})$
denote the solution of~\eqref{eq:kkt}.
For a general $\bm s_m$, only the equality right-hand side changes.
Standard multiparametric-QP sensitivity~\cite{bemporad2002} therefore gives an affine KKT solution while the active set remains valid. Its primal sensitivity to $\bm s_m$ is the first block $\bm D_{s,m}(t)$ satisfying
\begin{equation*}
\bm K_m(t)
\begin{bmatrix}
\bm D_{s,m}(t)\\ *\\ *
\end{bmatrix}
=
\begin{bmatrix}
\bm 0\\
-\bm F_{\mathcal O_m}\\
\bm 0
\end{bmatrix}.
\end{equation*}
The corresponding primal solution is
$\bm z_{\mathcal O_m}^0(\bm p_{\mathcal O_m})
+\bm D_{s,m}(t)[\bm s_m-\bm s_m^0(t)]$.
The critical region of $\mathcal A_m^0$ is the set of
$(\bm s_m,\bm p_{\mathcal O_m})$
for which this KKT solution is primal feasible and its selected inequality multipliers are nonnegative~\cite{bemporad2002}.

\subsection{Approximation and online update}

The previous subsection gives the local dependence of the remaining-network optimum on the exchanged flows. We now convert it into an approximation of $V_m$ and show how current information enters online. Define
\begin{equation*}
\Delta V_m
=
V_m(\bm s_m;\bm p_{\mathcal O_m})
-
V_m(\bm s_m^0(t);\bm p_{\mathcal O_m}),
\end{equation*}
where $\Delta\bm s_m=\bm s_m-\bm s_m^0(t)$. The subtracted term is constant in $(\bm z_m,\bm s_m)$ and therefore does not change the optimizer of~\eqref{eq:reduction}. Within the critical region, substituting the affine KKT solution into~\eqref{eq:value} gives
\begin{equation}\label{eq:approx}
\begin{aligned}
\widehat{\Delta V}_m
(\bm s_m;\bm p_{\mathcal O_m})
={}&
\bm v_m(\bm p_{\mathcal O_m})^\top
\Delta\bm s_m\\
&+
\tfrac12
\Delta\bm s_m^\top
\bm Q_m(t)
\Delta\bm s_m,
\end{aligned}
\end{equation}
where
$\bm v_m(\bm p_{\mathcal O_m})
=\bm F_{\mathcal O_m}^{\top}
\bm\pi_m^0(\bm p_{\mathcal O_m})$
is the marginal cost of the boundary exchanges and
$\bm Q_m(t)
=\bm D_{s,m}(t)^\top
\bm H_{\mathcal O_m}
\bm D_{s,m}(t)\succeq\bm 0$
is its curvature with respect to $\bm s_m$, that is, the Hessian of $V_m$ in the critical region.

Since the KKT right-hand side in~\eqref{eq:kkt} is affine in
$\bm p_{\mathcal O_m}$, the marginal-cost vector is affine as well:
\begin{equation}\label{eq:v-expansion}
\bm v_m(\bm p_{\mathcal O_m})
=
\bar{\bm v}_m(t)
+
\sum_{j\in\mathcal O_m}
\bm M_{mj}(t)
\bigl[
\bm p_j-\bar{\bm p}_j(t)
\bigr],
\end{equation}
where
$\bar{\bm v}_m(t)
=\bm v_m(\bar{\bm p}_{\mathcal O_m}(t))$.
Let $\bm B_{m,j}$ be the Jacobian of the right-hand side of~\eqref{eq:kkt} with respect to $\bm p_j$, and let
$\bm B_{s,m}=\col(\bm 0,\bm F_{\mathcal O_m},\bm 0)$.
Then
$\bm M_{mj}(t)
=\bm B_{s,m}^{\top}\bm K_m(t)^{-1}\bm B_{m,j}$.
Thus $\bm M_{mj}$ maps a deviation of the data at intersection $j$ to the resulting change in the boundary marginal cost. These quantities are computed offline.

At update $t$, IR-MPC uses the current data of $\mathcal R_m$ and the historical references elsewhere. Accordingly,
\begin{equation}\label{eq:slope}
\bm v_m^{\mathcal R_m}(t)
=
\bar{\bm v}_m(t)
+
\sum_{j\in\mathcal R_m\setminus\{m\}}
\bm M_{mj}(t)
\bigl[
\bm p_j(t)-\bar{\bm p}_j(t)
\bigr].
\end{equation}
Intersection $m$ is absent from the sum because~\eqref{eq:value} depends only on
$\bm p_{\mathcal O_m}$.
Using~\eqref{eq:slope} in~\eqref{eq:approx} gives the explicit local QP
\begin{equation}\label{eq:local}
\begin{aligned}
\min_{\bm z_m,\bm s_m}\quad&
J_m(\bm z_m;\bm p_m(t))
+
\bm v_m^{\mathcal R_m}(t)^\top\Delta\bm s_m\\
&+
\tfrac12
\Delta\bm s_m^\top
\bm Q_m(t)
\Delta\bm s_m\\
\mathrm{s.t.}\quad&
(\bm z_m,\bm s_m)\in
\mathcal X_m(\bm p_m(t)),\\
&
\bm s_m\in\mathcal S_m(t).
\end{aligned}
\end{equation}
This is~\eqref{eq:irmpc} with $\widehat V_m$ given by~\eqref{eq:approx}. Reading one additional intersection changes only the slope in~\eqref{eq:slope}; it does not enlarge the online optimization problem.

\section{Error Bounds and Information Regions}\label{sec:bounds}

Section~\ref{sec:irmpc} introduced two approximations relative to centralized MPC: the exact boundary value function is represented locally by a quadratic model, and current data are read only from $\mathcal R_m$. This section separates the approximation error into two parts, bounds their effect on the first control input, and then uses the resulting coupling measure to choose the information region.

\subsection{Error of the approximate boundary value function}\label{sec:boundary-error}

We first bound the error of $\widehat{\Delta V}_m$ itself. Let
$\widehat{\Delta V}_m^{\mathcal R_m}$
denote~\eqref{eq:approx} evaluated with~\eqref{eq:slope}, and let
$\widehat{\Delta V}_m^{\mathcal V}$
denote the same quadratic model with current data from all intersections.
The former is used by IR-MPC; the latter isolates the error of the quadratic approximation from the error caused by the unread data.

Throughout, $\|\cdot\|_2$ denotes the Euclidean norm and its induced matrix norm, and $\|\cdot\|_\infty$ the maximum norm. We measure the ranges of the data and of the exchanges by
\[
\begin{aligned}
b_j
&\ge
\max_{\bm p_j\in\Omega_j}
\norm{\bm p_j-\bar{\bm p}_j(t)}_2,\\
\rho_m(t)
&=
\max_{\bm s_m\in\mathcal S_m(t)}
\norm{\bm s_m-\bm s_m^0(t)}_2,\\
\omega_{mj}(t)
&=
\rho_m(t)b_j
\norm{\bm M_{mj}(t)}_2.
\end{aligned}
\]
and define
\begin{equation}\label{eq:delta}
\delta_m(\mathcal R_m;t)
=
\tfrac12
\norm{\bm Q_m(t)}_2
\rho_m(t)^2
+
\sum_{j\notin\mathcal R_m}
\omega_{mj}(t).
\end{equation}
The first term bounds the error of the quadratic approximation, and the second bounds the error caused by the unread data, in which $\omega_{mj}$ is the contribution of intersection $j$.

\begin{assumption}\label{as:validity}
For every $\bm p\in\Omega$,
$(\bm s_m^0(t),\bm p_{\mathcal O_m})$
lies in the critical region of $\mathcal A_m^0$, and the affine KKT primal solution
\begin{equation*}
\bm z_{\mathcal O_m}^0(\bm p_{\mathcal O_m})
+
\bm D_{s,m}(t)
[\bm s_m-\bm s_m^0(t)]
\end{equation*}
is feasible for~\eqref{eq:value} for every
$\bm s_m\in\mathcal S_m(t)$.
\end{assumption}

\begin{theorem}\label{thm:boundary}
Under Assumption~\ref{as:validity}, for every
$(\bm s_m,\bm p)\in\mathcal S_m(t)\times\Omega$,
\begin{equation}\label{eq:boundarybound}
\left|
\widehat{\Delta V}_m^{\mathcal R_m}
-
\Delta V_m
\right|
\le
\delta_m(\mathcal R_m;t).
\end{equation}
With full information,
\[
0\le
\widehat{\Delta V}_m^{\mathcal V}
-
\Delta V_m
\le
\delta_m(\mathcal V;t).
\]
Moreover, if
$m\in\mathcal R_m\subseteq\mathcal R'_m\subseteq\mathcal V$, then
\begin{equation}\label{eq:monotone}
\delta_m(\mathcal R_m;t)
-
\delta_m(\mathcal R'_m;t)
=
\sum_{j\in\mathcal R'_m\setminus\mathcal R_m}
\omega_{mj}(t)
\ge0.
\end{equation}
\end{theorem}

\proofhead
At $\bm s_m^0(t)$, Assumption~\ref{as:validity} makes the KKT solution optimal for every $\bm p\in\Omega$. Convexity therefore gives the supporting-hyperplane lower bound on $\Delta V_m$. The affine KKT primal solution is feasible for every $\bm s_m\in\mathcal S_m(t)$, so its cost gives an upper bound whose excess is at most
$\tfrac12\norm{\bm Q_m(t)}_2\rho_m(t)^2$.
By~\eqref{eq:v-expansion}, replacing current data outside $\mathcal R_m$ by their historical references changes only the linear term and contributes at most
$\sum_{j\notin\mathcal R_m}\omega_{mj}(t)$.
Combining the two terms proves~\eqref{eq:boundarybound}, and~\eqref{eq:monotone} follows directly from~\eqref{eq:delta}.\qedsq

Theorem~\ref{thm:boundary} separates the error that remains even with full information from the additional error introduced by restricting the online information region.

\subsection{First-control-input error}\label{sec:input-error}

We now translate this error into an error of the first control input. Define the hybrid data vector $\widetilde{\bm p}$ by using the current data $\bm p_j(t)$ for $j\in\mathcal R_m$ and the historical reference $\bar{\bm p}_j(t)$ otherwise. At $\widetilde{\bm p}$, the local QP used by intersection $m$ is the full-information version of~\eqref{eq:local}. Therefore,
\begin{equation}\label{eq:input-split}
\begin{aligned}
&\norm{\bm a_m^{\mathcal R_m}(t)-\bm a_m^\star(\bm p(t))}_\infty\\
&\le\underbrace{\norm{\bm a_m^{\mathcal R_m}(t)-\bm a_m^\star(\widetilde{\bm p})}_\infty}_{\zeta_{m,1}}
+\underbrace{\norm{\bm a_m^\star(\widetilde{\bm p})-\bm a_m^\star(\bm p(t))}_\infty}_{\zeta_{m,2}}.
\end{aligned}
\end{equation}
The first term is caused by the approximation of $V_m$; the second is the change in the centralized first control input caused by the unread data.

To bound $\zeta_{m,1}$, let $\kappa_m>0$ satisfy the curvature condition
\begin{equation}\label{eq:curvature}
\Delta\bm z^\top\bm H\Delta\bm z
\ge
\kappa_m
\norm{(\Delta\bm z)_{\mathcal I_m}}_2^2,
\qquad
\bm E\Delta\bm z=\bm 0.
\end{equation}
For an updating intersection, the green-time regularization~\eqref{eq:cost-green} includes the difference between the first optimized cycle and the fixed preceding cycle, and therefore provides $\kappa_m\ge\mu_g>0$ for the implemented green-time block.

Let $\bm a_m^{\mathcal V}(\bm p)$ denote the first control input of~\eqref{eq:local} when all current data are used.

\begin{lemma}\label{lem:full-information}
Under Assumptions~\ref{as:coverage}--\ref{as:validity} and~\eqref{eq:curvature}, for every $\bm p\in\Omega$,
\begin{equation}\label{eq:full-info-input}
\norm{
\bm a_m^{\mathcal V}(\bm p)
-\bm a_m^\star(\bm p)
}_\infty
\le
\sqrt{2\delta_m(\mathcal V;t)/\kappa_m}.
\end{equation}
\end{lemma}

\proofhead
By Theorem~\ref{thm:boundary}, the full-information approximate reduced objective exceeds the exact reduced objective by at most $\delta_m(\mathcal V;t)$. Assumption~\ref{as:coverage} places a centralized optimizer inside $\mathcal S_m(t)$, while Assumption~\ref{as:validity} ensures that the approximate local optimizer can be completed by a feasible remaining-network solution. Hence the corresponding feasible centralized point has objective gap at most $\delta_m(\mathcal V;t)$. Convex-QP optimality~\cite[Sec.~4.2.3]{boyd2004} together with~\eqref{eq:curvature} gives~\eqref{eq:full-info-input}.\qedsq

Since
$\delta_m(\mathcal V;t)
=\tfrac12\norm{\bm Q_m(t)}_2\rho_m(t)^2$,
the full-information component depends only on the error of the quadratic approximation.

To bound $\zeta_{m,2}$, we use the standard piecewise-affine sensitivity of a multiparametric QP~\cite{bemporad2002}. Assume that
$\bm a_m^\star$ is continuous on $\Omega$ and affine on each member $\alpha$ of a finite polyhedral cover of $\Omega$. Let $\bm\Theta_{mj}^{\alpha}$ denote the coefficient of $\bm p_j$ in that affine map. The coupling bound is
\begin{equation*}
W_{mj}^{\mathrm{ub}}
=
b_j
\max_{\alpha}
\norm{\bm\Theta_{mj}^{\alpha}}_2,
\qquad
j\in\mathcal O_m.
\end{equation*}
It bounds the change in seconds of green time that the admissible variation of the data at $j$ can cause in the centralized first control input of $m$.

For a tolerance $\varepsilon_m\ge0$ on the deviation of the first control input from that of centralized MPC, the following theorem holds.

\begin{theorem}\label{thm:selection}
Under Assumptions~\ref{as:coverage}--\ref{as:validity},~\eqref{eq:curvature}, and the piecewise affinity of $\bm a_m^\star$, any information region $\mathcal R_m\ni m$ satisfying
\begin{equation}\label{eq:coupling-tolerance}
\sum_{j\notin\mathcal R_m}
W_{mj}^{\mathrm{ub}}
\le
\varepsilon_m
-
\sqrt{2\delta_m(\mathcal V;t)/\kappa_m}
\end{equation}
guarantees
\[
\bm a_m^{\mathcal R_m}(t)=\bm a_m^\star(\bm p(t))+\bm\xi_m(t),\qquad
\norm{\bm\xi_m(t)}_\infty\le\varepsilon_m
\]
for $\bm p(t)\in\Omega$, with the first-control-input disturbance $\bm\xi_m(t)$.
\end{theorem}

\proofhead
Lemma~\ref{lem:full-information}, applied at $\widetilde{\bm p}$, bounds $\zeta_{m,1}$. The segment from $\widetilde{\bm p}$ to $\bm p(t)$ lies in $\Omega$ because $\Omega$ is a product set, and only the data of intersections outside $\mathcal R_m$ change along it. Splitting this segment across the affine regions gives
$\zeta_{m,2}
\le\sum_{j\notin\mathcal R_m}W_{mj}^{\mathrm{ub}}$.
Equation~\eqref{eq:input-split} and~\eqref{eq:coupling-tolerance} then give the result.\qedsq

Theorem~\ref{thm:selection} gives the design principle: current data may be omitted when their combined influence on the centralized first control input fits within the remaining approximation tolerance.

\subsection{Information regions and the IR-MPC algorithm}\label{sec:coupling}

The worst-case bound above provides the theoretical criterion, whereas the implementation estimates the couplings from historical data. Each historical day provides a sample $q$ at every update time $t$, with network data $\bm p^{(q)}$. At each sample, \eqref{eq:full} is solved with $\bm p^{(q)}$. For intersection $j$, the numbers of vehicles, entry queues, and arrival rates are then shifted by $\pm\sigma_\ell(t)$, with negative values truncated at zero, and~\eqref{eq:full} is re-solved using the resulting data $\bm p^{(q,j,\pm)}$. Here $\sigma_\ell(t)$ is the historical standard deviation of the corresponding entry $\ell$ at time $t$. The empirical coupling is
\begin{equation}\label{eq:coupling}
\begin{aligned}
W_{mj}(q)
&=
\max_{\pm}
\norm{
\bm a_m^\star(\bm p^{(q,j,\pm)})
-
\bm a_m^\star(\bm p^{(q)})
}_\infty,\\
\bar W_{mj}(\tau)
&=
\frac{1}{|\mathcal H_{m,\tau}|}
\sum_{q\in\mathcal H_{m,\tau}}
W_{mj}(q),
\end{aligned}
\end{equation}
where $\mathcal H_{m,\tau}$ contains the samples of demand period $\tau$ at which a cycle of $m$ starts. Thus $\bar W_{mj}(\tau)$ measures the typical sensitivity of the centralized first control input of $m$ to the data of $j$ under the historical traffic pattern.

For practical selection, we use nested candidate regions. Let
$\mathcal R_{m,i}$ contain $m$ and its $i-1$ nearest other intersections in hop distance, and let $\mathcal N$ be the candidate set of region sizes with
$1,|\mathcal V|\in\mathcal N$.
Replacing the worst-case coupling in~\eqref{eq:coupling-tolerance} by the historical mean coupling and the right-hand side by a tolerance $\varepsilon_W$ on the change of the first control input caused by the unread data, the selected size is
\begin{equation}\label{eq:read-size}
i_m(\tau)
=
\min\left\{
i\in\mathcal N:
\sum_{j\notin\mathcal R_{m,i}}
\bar W_{mj}(\tau)
\le
\varepsilon_W
\right\},
\end{equation}
and $\mathcal R_m(\tau)=\mathcal R_{m,i_m(\tau)}$.
Thus network distance defines the nested candidate sets, while the measured couplings determine how far the region must extend.

\begin{algorithm}[!t]
\caption{IR-MPC}\label{alg:irmpc}
\begin{algorithmic}[1]
\Require historical data, demand periods, $\mathcal N$, $\varepsilon_W$
\Statex Offline:
\State for each $t$ and $m$, compute $\bm s_m^0(t)$, $\mathcal S_m(t)$, $\bm Q_m(t)$, $\bar{\bm v}_m(t)$, $\bm M_{mj}(t)$ (Sec.~\ref{sec:irmpc})\label{alg:l-bf}
\State compute $\bar W_{mj}(\tau)$ by~\eqref{eq:coupling} and $\mathcal R_m(\tau)$ by~\eqref{eq:read-size}\label{alg:l-c}
\Statex Online, at each update $t$:
\State each intersection $m$ whose cycle starts reads $\bm p_j(t)$ for $j\in\mathcal R_m(\tau)$
\State compute $\bm v_m^{\mathcal R_m}(t)$ by~\eqref{eq:slope}\label{alg:l-read}
\State solve~\eqref{eq:local} and implement $\bm a_m^{\mathcal R_m}(t)$\label{alg:l-solve}
\end{algorithmic}
\end{algorithm}

Algorithm~\ref{alg:irmpc} separates the historical work from the online controller: $\widehat V_m$ and the information regions are prepared offline, while each online update uses the selected current data and solves the local QP once.

\section{Simulation Study}\label{sec:experiments}

This section evaluates IR-MPC from mechanism to network-scale performance. We first specify the experimental setup, then examine whether the couplings identify compact but potentially nonlocal information regions. We next compare the approximation of the first control input and the closed-loop traffic performance with centralized MPC and iterative OSD, and finally test how computation and information requirements scale with network size.

\begin{figure}[H]
\centering
\includegraphics[width=\columnwidth]{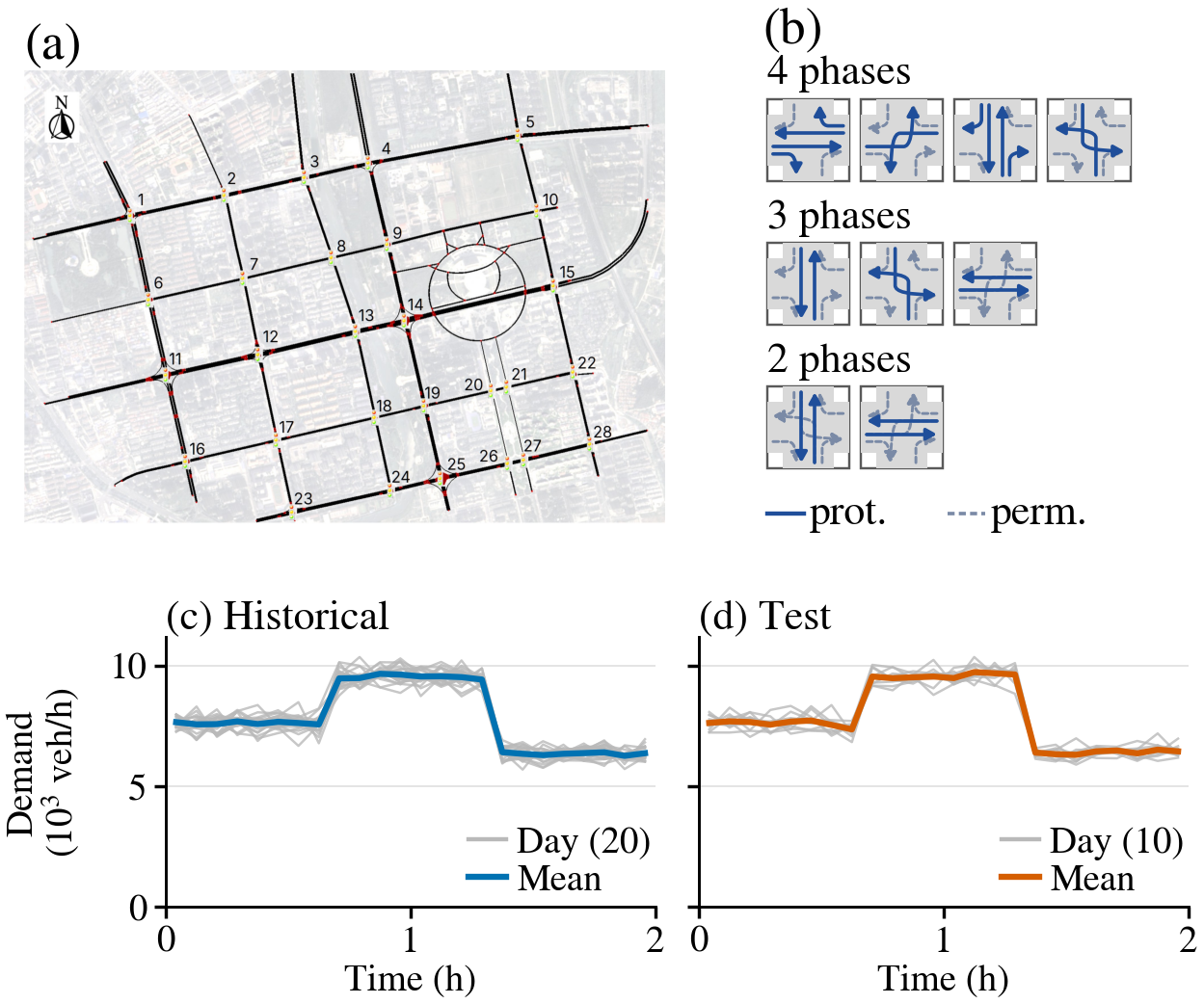}
\caption{Test network, phase combinations, and departure rates on the historical and test days.}
\label{fig:scenario}
\end{figure}

\subsection{Simulation setup}\label{sec:setup}

The experiments are conducted in SUMO~\cite{sumo2018} on a real 28-intersection network in Lianyungang, China~\cite{tan2026privatemp,tan2026regret} (as shown in Fig.~\ref{fig:scenario}). Traffic demand is generated by Poisson arrivals on 36 origin flows over three 2,400-s demand periods with multipliers of 1.2, 1.5, and 1.0, followed by a clearance period~\cite{tan2026regret}. Twenty independently generated days are used as historical data and ten additional days for testing.

The network contains 21 intersections with 60-s cycles and 7 intersections with 120-s cycles. The MPC uses a 60-s prediction step and an eight-step horizon. Saturation flows are set to 1,800~veh/h/lane unless measured values are available, and turning ratios and historical arrival rates are estimated from the historical days. Entry arrival rates are estimated from the most recent 300~s of observations. The objective weights are $\mu_g=10^{-4}$ and $\mu_R=0.01$. Information regions are selected using $\varepsilon_W=1$~s, while the admissible boundary-flow sets are constructed from historical values with a 10\% margin. All QPs are solved using Clarabel~0.11.1~\cite{goulart2024clarabel} on a laptop with an AMD Ryzen~7 5825U CPU and 16~GB RAM.

We compare IR-MPC with centralized MPC, which solves the full network-level QP~\eqref{eq:full}, and with OSD~\cite{shin2020schwarz} with one local subproblem per intersection and a one-hop or two-hop overlap (OSD-1, OSD-2), or with a one-hop overlap on four seven-intersection subnetworks obtained by recursive bisection weighted by the estimated inter-intersection couplings (P-OSD). Each OSD variant is evaluated with different iteration counts $K$ and initialized from the centralized solution at the historical reference data $\bar{\bm p}(t)$.

\subsection{Inter-intersection coupling and information regions}
\label{sec:coupling-results}

\begin{figure}[!b]
\centering
\includegraphics[width=\columnwidth]{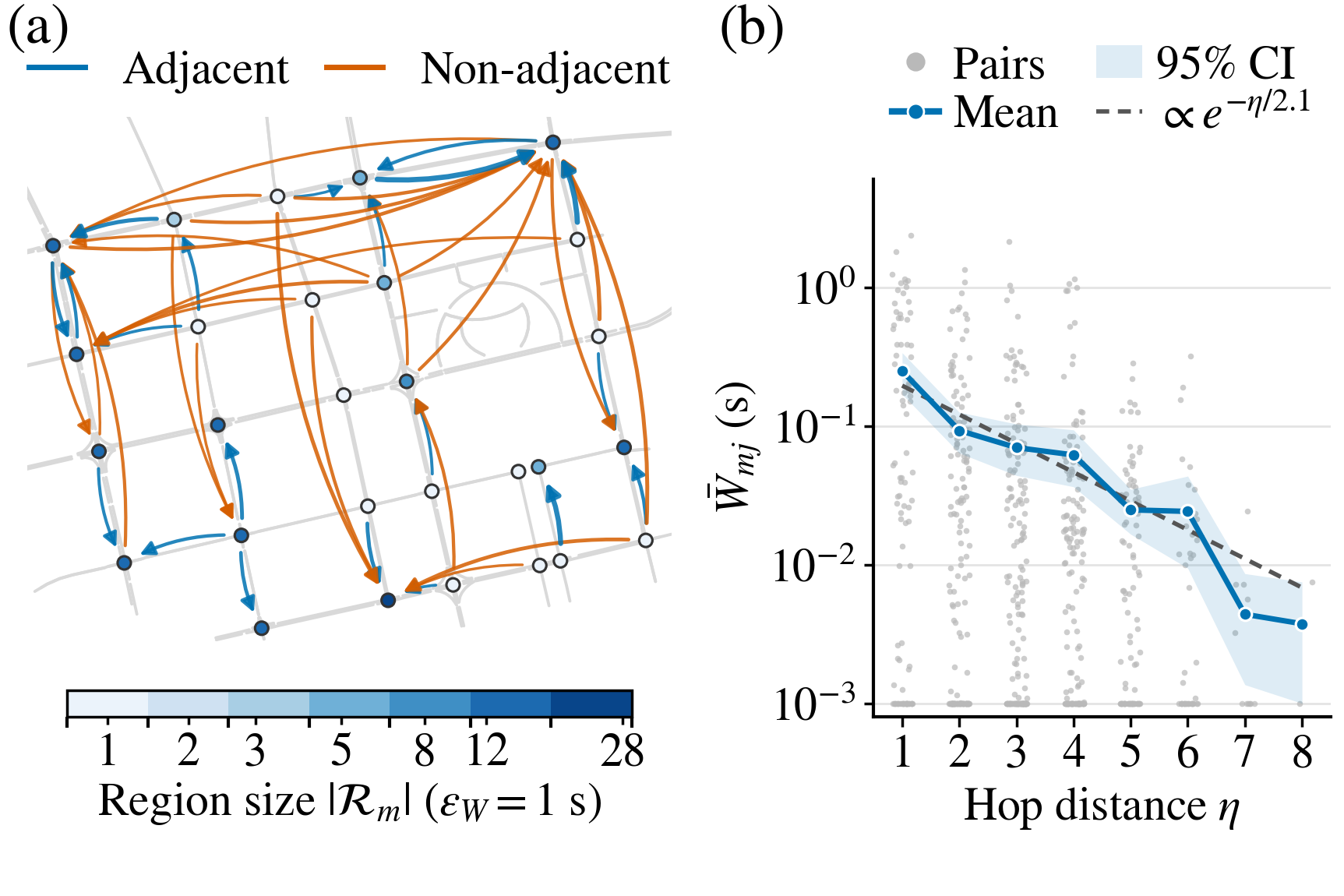}
\caption{Spatial structure of inter-intersection coupling: (a) strongly coupled pairs during the high-demand period; (b) coupling strength versus hop distance.}
\label{fig:historical-coupling}
\end{figure}

Figure~\ref{fig:historical-coupling} reveals a clear but imperfect spatial decay of inter-intersection coupling. The average coupling decreases approximately exponentially with hop distance ($R^2=0.89$), consistent with sensitivity decay in graph-structured optimization~\cite{shin2022}. However, network proximity alone does not identify the intersections that matter most: more than half of the pairs with $\bar W_{mj}\ge0.25$~s are nonadjacent. Thus, hop distance provides a useful spatial prior, but not a reliable criterion for determining the information required to reproduce the centralized first control input.

This distinction is reflected in the selected information regions. Under~\eqref{eq:read-size}, the regions expand during periods with active demand and contract during network clearance, indicating that the spatial extent of relevant information varies with traffic conditions rather than remaining fixed by network geometry. Despite the nonlocal couplings, the resulting regions remain small relative to the 28-intersection network, and 46\% of the intersections require only their own current data throughout the experiment. These results support the central premise of IR-MPC: the centralized first control input can depend on distant traffic states, but only a limited subset of them needs to be updated online.

\subsection{Solution approximation and computational efficiency}
\label{sec:vs-distributed}

Table~\ref{tab:distributed} compares the methods on 50 test samples. The control-input error is the maximum phase-wise difference between the next-cycle green times produced by a method and those of centralized MPC at the same traffic state. The computation time is the median, over the 50 test samples, of the time of the slowest simultaneously solved local QP or subnetwork, summed over the iterations for OSD; it assumes one processor per local problem and excludes communication time.

\begin{table}[h!]
\centering
\caption{Solution approximation and computation time.}
\label{tab:distributed}
\begingroup
\footnotesize
\setlength{\tabcolsep}{1.5pt}
\renewcommand{\arraystretch}{1.05}
\resizebox{\columnwidth}{!}{%
\begin{tabular}{@{}lrrrrr@{}}
\toprule
Method & \shortstack{Solved\\(\%)} & \shortstack{Mean\\error (s)} &
\shortstack{Error\\$\le1$ s (\%)} &
\shortstack{Error\\$\le2$ s (\%)} &
\shortstack{Comp.\\time (s)} \\
\midrule
Centralized MPC & -- & -- & -- & -- & 0.97 \\
IR-MPC & 99.5 & 0.155 & 96.1 & 99.8 & 0.11 \\
OSD-1, $K=1$ & 100.0 & 0.285 & 90.2 & 96.5 & 0.18 \\
OSD-1, $K=4$ & 99.6 & 0.074 & 98.0 & 99.8 & 0.71 \\
OSD-2, $K=1$ & 99.6 & 0.236 & 92.7 & 96.8 & 0.40 \\
OSD-2, $K=4$ & 98.7 & 0.015 & 100.0 & 100.0 & 1.61 \\
P-OSD, $K=1$ & 100.0 & 0.217 & 93.3 & 97.5 & 0.38 \\
P-OSD, $K=4$ & 97.0 & 0.029 & 99.6 & 100.0 & 1.53 \\
\bottomrule
\end{tabular}}
\endgroup
\end{table}

With only one OSD iteration, enlarging the local optimization problem through overlap or partitioning does not fully recover the effect of the rest of the network: all one-iteration variants produce larger control errors than IR-MPC. In contrast, IR-MPC incorporates the effect of the external network through $\widehat V_m$ before the online optimization is solved.

Additional OSD iterations can recover or exceed the accuracy of IR-MPC, but this comes at a substantial computational cost. At four iterations, the best OSD configurations approach the centralized solution closely, yet require 6--15 times the computation time of IR-MPC and, for the larger overlaps, take roughly as long as or longer than centralized MPC itself. The comparison therefore exposes the trade-off that IR-MPC is designed to avoid: iterative decomposition recovers network-level coordination through repeated online exchange, whereas IR-MPC shifts most of this coordination effort offline and retains only one local solve online.

\subsection{Closed-loop traffic performance}
\label{sec:vs-central}

We then ask whether the approximation of the first control input remains reliable when applied repeatedly in feedback. This is important because small one-step differences could, in principle, alter subsequent traffic states. Table~\ref{tab:summary} compares the methods over the ten test days (mean $\pm$ 95\% confidence interval). Delay is SUMO time loss~\cite{tan2026regret}, and the total time spent (TTS) includes both trip duration and departure delay. The closed-loop cost $J_{\mathrm{cl}}$ evaluates the objective~\eqref{eq:ocp-obj} along the simulated trajectory of each test day, and $\Delta J_{\mathrm{cl}}$ is its relative difference from centralized MPC.

\begin{table}[h!]
\centering
\caption{Closed-loop traffic performance and cost.}
\label{tab:summary}
\begingroup
\footnotesize
\setlength{\tabcolsep}{1.5pt}
\renewcommand{\arraystretch}{1.05}
\resizebox{\columnwidth}{!}{%
\begin{tabular}{@{}lrrrr@{}}
\toprule
Method &
\shortstack{Mean delay\\(s/veh)} &
\shortstack{TTS\\(veh$\cdot$h)} &
\shortstack{$\Delta$ delay\\(s/veh)} &
\shortstack{$\Delta J_{\mathrm{cl}}$\\(\%)} \\
\midrule
Centralized MPC & $130.45\pm2.08$ & $1465.2\pm13.8$ & -- & -- \\
IR-MPC & $130.49\pm2.26$ & $1465.4\pm14.6$ & $+0.04\pm0.46$ & $+0.02\pm0.11$ \\
OSD-1, $K=1$ & $134.45\pm8.34$ & $1482.6\pm36.4$ & $+4.00\pm8.67$ & $+1.49\pm3.18$ \\
OSD-1, $K=4$ & $161.83\pm62.85$ & $1601.8\pm272.6$ & $+31.38\pm62.83$ & $+11.73\pm23.46$ \\
OSD-2, $K=1$ & $183.35\pm89.76$ & $1698.5\pm396.0$ & $+52.90\pm89.83$ & $+20.83\pm35.96$ \\
OSD-2, $K=4$ & $131.00\pm3.20$ & $1467.6\pm18.7$ & $+0.54\pm1.27$ & $+0.12\pm0.35$ \\
P-OSD, $K=1$ & $146.66\pm36.35$ & $1536.3\pm158.9$ & $+16.20\pm36.22$ & $+6.05\pm13.37$ \\
P-OSD, $K=4$ & $131.88\pm3.02$ & $1471.5\pm16.9$ & $+1.43\pm2.60$ & $+0.32\pm0.59$ \\
\bottomrule
\end{tabular}}
\endgroup
\end{table}

At every IR-MPC update, we also solve centralized MPC from the same traffic state without applying its solution, and compare the two control inputs. Although the IR-MPC control inputs are approximations, 85\% coincide exactly with those of centralized MPC at the implemented one-second resolution and 99.3\% remain within 2~s. These small discrepancies do not accumulate: the number of vehicles in the network stays within 0.24\% of centralized MPC, and the closed-loop cost differs by only 0.02\%. The mean-delay difference is only 0.03\%, with no significant difference at the 95\% level in any demand period. This indicates that reproducing the centralized first control input to high, rather than exact, numerical precision is sufficient to preserve its closed-loop traffic performance in this network.

In contrast, some control inputs of four of the six OSD variants deviate considerably from the centralized solution on a few days, which leads to severe delays on those days. Their mean delay is therefore much larger than that of IR-MPC. This shows that consistently accurate control inputs are essential for closed-loop performance, and IR-MPC provides them without iterative coordination within a control update.

Only 0.45\% of IR-MPC local updates are infeasible or unsolved; the controller then retains the previous green times, and a failed OSD subproblem keeps its previous iterate. Most such cases are associated with insufficient coverage of the admissible boundary-flow set, identifying boundary-set construction as the main practical limitation in these experiments.

\subsection{Scalability on synthetic grid networks}\label{sec:scale}

Synthetic grids with 49, 100, 200, and 500 intersections are generated with heterogeneous link lengths and partially removed minor links. Because microscopic simulation becomes prohibitively expensive at this scale, we use a link-level traffic-flow model with the same MPC structure as in Section~\ref{sec:setup}. Four historical days are used to construct the information regions and $\widehat V_m$, and two additional days are used for testing. All comparisons are made at a peak-demand control update. The reported IR-MPC time is the median computation time of the slowest local QP when one processor is assigned to each intersection.

\begin{table}[h!]
\centering
\caption{Scalability of centralized MPC and IR-MPC.}
\label{tab:scale}
\begingroup
\footnotesize
\setlength{\tabcolsep}{4pt}
\renewcommand{\arraystretch}{1.05}
\begin{tabular}{@{}rrrrr@{}}
\toprule
\shortstack{Inter-\\sections} &
\shortstack{Centralized\\time (s)} &
\shortstack{IR-MPC\\time (s)} &
\shortstack{Mean\\$|\mathcal R_m|$} &
\shortstack{Error $\le2$ s\\(\%)} \\
\midrule
49  & 0.63 & 0.016 & 2.0 & 100.0 \\
100 & 2.04 & 0.016 & 2.3 & 98.0 \\
200 & 5.36 & 0.016 & 1.7 & 99.5 \\
500 & 21.9 & 0.024 & 1.8 & 100.0 \\
\bottomrule
\end{tabular}
\endgroup
\end{table}

Table~\ref{tab:scale} shows a fundamentally different scaling behavior between the two control architectures. The computation time of centralized MPC increases rapidly with network size because the full network is optimized jointly. In contrast, the online computation time of IR-MPC remains nearly unchanged even when the network grows by an order of magnitude. This occurs because each local controller optimizes only its own variables, while the effect of the remaining network is represented through $\widehat V_m$ rather than additional online variables.

The mean information-region size remains close to two intersections across all network sizes, indicating that enlarging the network does not require each controller to observe a proportionally larger portion of it. At the same time, 98--100\% of the test control inputs remain within 2~s of centralized MPC. The scalability of IR-MPC therefore comes from two properties acting together: the size of each local optimization problem remains essentially fixed, and the amount of current network information required for an accurate centralized approximation remains sparse as the network grows.

\section{Conclusion}

This paper developed IR-MPC, which approximates the network-level behavior of centralized MPC with high accuracy from a small information region, without solving or iteratively coordinating a network-wide problem online. A boundary value function represents the influence of the remaining network on each intersection. On the real 28-intersection network, IR-MPC keeps 99.3\% of its control inputs within 2~s of centralized MPC while changing mean delay by only 0.03\%. This indicates that a high-fidelity approximation of the centralized first control input is sufficient to preserve its closed-loop traffic performance. The scalability experiments show the potential of IR-MPC for large networks: in networks of up to 500 intersections, the mean information region contains only about two intersections, and the control approximation remains highly accurate. Future work will investigate online adaptation under changing traffic conditions and establish closed-loop stability guarantees for the resulting non-iterative distributed controller.

\section*{Acknowledgment}
The authors used Claude to polish parts of the text and generate code; the authors reviewed all content and take full responsibility for it.

\bibliographystyle{IEEEtran}
\bibliography{references}
\end{document}